%% file: main.tex
\documentclass[letterpaper,twocolumn,10pt]{article}
\usepackage{usenix}

\usepackage{tikz}
\usepackage{amsmath}

\usepackage{xspace}
\usepackage{listings}
\usepackage{pgf}
\usepackage{tikz}
\usetikzlibrary{arrows,automata}
\usetikzlibrary{arrows.meta}
\usetikzlibrary{shadows}
\usetikzlibrary{calc}
\usetikzlibrary{positioning}
\usetikzlibrary{fit}
\usepackage[vlined,ruled,linesnumbered]{algorithm2e}
\usepackage[noend]{algpseudocode}
\usepackage{amssymb}
\usepackage[labelfont=bf,textfont=bf]{caption}
\usepackage{subcaption}
\usepackage{multirow}
\usepackage{hyperref}
\usepackage{pgfplots}
\usepackage{pgfplotstable}
\usepackage{microtype}
\usepackage{balance}
\usepackage{tabularx}
\usepackage{xcolor}
\usepackage{colortbl}
\usepackage{enumitem}
\usepackage[norndcorners,customcolors,nofill]{hf-tikz}
\usepackage{xstring}
\usepackage{adjustbox}
\usepackage{tabularx,booktabs}
\usepackage{array}
\usepackage{graphicx}
\usepackage{makecell}

\usepackage{rust}
\newcommand\circom{\textsc{Circom}\xspace}
\newcommand\noir{\textsc{Noir}\xspace}
\newcommand\corset{\textsc{Corset}\xspace}
\newcommand\gnark{\textsc{Gnark}\xspace}
\newcommand\circuzz{\textsc{Circuzz}\xspace}
\newcommand\arguzz{\textsc{Arguzz}\xspace}
\newcommand\il{\textsc{CircIL}\xspace}

\newcommand\tool{\textsc{Liezz}\xspace}

\newcommand{\positive}[1]{\textcolor{evalgreen}{#1}}
\newcommand{\negative}[1]{\textcolor{evalred}{#1}}

\newcommand\code[1]{\lstinline[style=basic]{#1}}
\newcommand\codeil[1]{\lstinline[style=il]{#1}}

\newsavebox\slantboxbox
\newcommand{\slantbox}[2][0.22]{%
  \mbox{%
    \sbox{\slantboxbox}{#2}%
    \hskip\wd\slantboxbox
    \pdfsave
    \pdfsetmatrix{1 0 #1 1}%
    \llap{\usebox{\slantboxbox}}%
    \pdfrestore
  }%
}

\newcommand\bug[1]{%
  \IfEqCase{#1}{%
    {1}{#1}% circom flipsign
    {2}{\href{https://github.com/LFDT-Lineth/zkc/issues/955}{#1}}% corset MIR: Dave's internal issue
    {3}{\href{https://github.com/LFDT-Lineth/zkc/issues/1586}{#1}}% corset AIR
    {4}{\href{https://github.com/LFDT-Lineth/zkc/issues/1464}{#1}}% corset limb width: Dave's internal issue
    {5}{\href{https://github.com/LFDT-Lineth/zkc/issues/1505}{#1}}% corset small fields
    {6}{#1}% corset check_fail: confirmed via Signal, no public issue
    {7}{#1}% gnark twistededwards: reported via Telegram, no public issue
    {8}{#1}% gnark lookup: reported via Telegram, no public issue
    {9}{#1}% noir ite: reported via Signal, no public issue
    {10}{#1}% noir is_infinite
    {11}{#1}% noir write_vk: reported via Signal, no public issue
    {12}{\href{https://github.com/noir-lang/noir/issues/12034}{#1}}% noir vk scalar
    {13}{\href{https://github.com/noir-lang/noir/security/advisories/GHSA-683h-pgp9-8cq4}{#1}}% noir cast ignored
  }[\PackageError{bug}{Undefined option to bug: #1}{}]%
}

\newcommand\fix[1]{%
  \IfEqCase{#1}{%
    {A}{\href{https://github.com/iden3/snarkjs/pull/616}#1}% circom flipsign
    {B}{\href{https://github.com/LFDT-Lineth/zkc/pull/1587}#1}% corset air: fixed, internal issue/fix not public in notes
    {C}{\href{https://github.com/LFDT-Lineth/zkc/pull/1466}{#1}}% corset limb width
    {D}{\href{https://github.com/Consensys/gnark/pull/1751}{#1}}% gnark twistededwards
    {E}{\href{https://github.com/Consensys/gnark/pull/1754}{#1}}% gnark lookup
    {F}{\href{https://github.com/AztecProtocol/aztec-packages/pull/17998}{#1}}% noir ite
    {G}{\href{https://github.com/noir-lang/noir/pull/11925}#1}% noir is_infinite: note: there is a separate fix for bb: https://github.com/AztecProtocol/aztec-packages/pull/21865
    {H}{\href{https://github.com/noir-lang/noir/pull/11780/changes/a2219329482262e103d6be0780027826d96beecd}{#1}}% noir write_vk
    {I}{\href{https://github.com/noir-lang/noir/pull/12039}#1}% noir vk_scalar
    {J}{\href{https://github.com/noir-lang/noir/pull/12582}#1}% noir cast_ignored
  }[\PackageError{fix}{Undefined option to fix: #1}{}]%
}
\definecolor{gray}{rgb}{0.5, 0.5, 0.5}
\definecolor{light-gray}{gray}{0.77}
\definecolor{BrickRed}{rgb}{0.8, 0.25, 0.33}
\definecolor{Black}{rgb}{0.0, 0.0, 0.0}
\definecolor{DarkBlue}{rgb}{0.0, 0.0, 0.55}
\definecolor{Crimson}{rgb}{0.86, 0.08, 0.24}
\definecolor{SlateGrey}{rgb}{0.44, 0.5, 0.56}
\definecolor{lightorange}{HTML}{FFB74D}
\definecolor{blue}{rgb}{0.0, 0.0, 1.0}
\definecolor{magenta}{rgb}{0.79, 0.08, 0.48}
\definecolor{evalgreen}{RGB}{28,105,74}
\definecolor{evalred}{RGB}{145,38,50}
\lstdefinestyle{il}{%
  language         = C,%
  alsoletter       = -,%
  morekeywords     = [1]{inputs, outputs, assert},%
  morekeywords     = [2]{},%
  keywordstyle     = [2]\bfseries,%
  morekeywords     = [3]{},%
  keywordstyle     = [3]\bfseries,%
  keywordstyle     = \bfseries,%
  comment          = [l]{//},%
  commentstyle     = \ttfamily\color{Black!60}\small\lst@ifdisplaystyle\small\fi,%
  basicstyle       = \ttfamily\small\lst@ifdisplaystyle\small\fi,%
  emph             = {},%
  emphstyle        = {\color{teal}\bfseries},%
  stringstyle      = \color{BrickRed},%
  columns          = [c]fixed,%
  aboveskip        = 0mm,%
  belowskip        = 2mm,%
  keepspaces       = true,%
  mathescape       = true,%
  escapechar       = @,%
  tabsize          = 2,%
  numbers          = left,%
  numberstyle      = \tiny\color{Black!70},%
  numbersep        = 4pt,%
  stepnumber       = 1,%
  firstnumber      = 1,%
  showstringspaces = false,%
  captionpos       = b,%
  extendedchars    = true,%
  upquote          = true,%
  abovecaptionskip = 0mm,%
  belowcaptionskip = 0mm,%
  xleftmargin      = 3mm,%
  moredelim        = **[is][{\btHL[fill=light-gray]}]{°}{°},
  morecomment      = [s]{/*}{*/}
}

\lstdefinestyle{basic}{%
  language         = Rust,%
  alsoletter       = -,%
  morekeywords     = [1]{signal,input,output,public,template,component,var,function,return,if,else,for,while,do,log,assert,include,parallel,pragma,circom,custom_templates,defcolumns,defconstraint,let,if,pub,fn,new,type,func,nil},%
  morekeywords     = [2]{\~or!,neq!,eq!,is-not-zero!,vanishes!,if-not-zero,is-not-zero,is-zero,to\_be\_bytes,bytes32\_to\_field,SetString,AssertIsEqual,AssertIsLessOrEqual,SetUint64,Neg,Compiler,FieldBitLen},%
  keywordstyle     = [2]\color{teal}\bfseries,%
  morekeywords     = [3]{},%
  otherkeywords    = {macro_rules!},%
  keywordstyle     = [3]\color{BrickRed}\bfseries,%
  keywordstyle     = \bfseries\color{DarkBlue},%
  comment          = [l]{//},%
  commentstyle     = \ttfamily\color{Black!60}\small\lst@ifdisplaystyle\small\fi,%
  basicstyle       = \ttfamily\small\lst@ifdisplaystyle\small\fi,%
  emph             = {},%
  emphstyle        = {\color{teal}\bfseries},%
  stringstyle      = \color{BrickRed},%
  columns          = [c]fixed,%
  aboveskip        = 0mm,%
  belowskip        = 2mm,%
  keepspaces       = true,%
  mathescape       = true,%
  escapechar       = @,%
  tabsize          = 2,%
  numbers          = left,%
  numberstyle      = \tiny\color{Black!70},%
  numbersep        = 4pt,%
  stepnumber       = 1,%
  firstnumber      = 1,%
  showstringspaces = false,%
  captionpos       = b,%
  extendedchars    = true,%
  upquote          = true,%
  abovecaptionskip = 0mm,%
  belowcaptionskip = 0mm,%
  xleftmargin      = 3mm,%
  moredelim        = **[is][{\btHL[fill=light-gray]}]{°}{°},
  morecomment      = [s]{/*}{*/}
}

\begin{document}
%-------------------------------------------------------------------------------

\pagestyle{empty}

%don't want date printed
\date{}

\title{\Large \bf Lie to Me: Finding Bugs in ZK DSL Toolchains with Adversarial Witness Injection}

%\author{}
\author{
{\rm Sebastian Watzinger}\\
\href{mailto:sebastian.watzinger@tuwien.ac.at}{sebastian.watzinger@tuwien.ac.at}\\
TU Wien, Austria
\and
{\rm Christoph Hochrainer}\\
\href{mailto:christoph.hochrainer@tuwien.ac.at}{christoph.hochrainer@tuwien.ac.at}\\
TU Wien, Austria
\and
{\rm Valentin W\"ustholz}\\
\href{mailto:valentin.wustholz@diligence.security}{valentin.wustholz@diligence.security}\\
Consensys Diligence, Austria
\and
{\rm Maria Christakis}\\
\href{mailto:maria.christakis@tuwien.ac.at}{maria.christakis@tuwien.ac.at}\\
TU Wien, Austria
} % end author

\maketitle

\begin{abstract}
Zero-knowledge domain-specific language (ZK DSL) toolchains compile
programs into constraint systems and generate witnesses for
cryptographic proofs. Bugs in these toolchains can leave the enforced
constraints weaker than the source-program semantics, admitting proofs
for invalid executions. Such soundness bugs may remain invisible to
valid-execution testing because all valid executions still behave
correctly.

We present \tool, a testing framework that generates ZK DSL programs
and exposes these bugs through \emph{adversarial witness injection}.
For each generated deterministic program, \tool executes two public
input assignments with different outputs and splices their witnesses,
combining the input of one execution with the output of the other. The
resulting witness is invalid by construction. A correct toolchain must
reject it; acceptance exposes a soundness bug. Controlled divergence
and multiple witness-splicing strategies preserve enough consistency
to expose missing constraints. \tool also generates parameterized
standard-library calls to reach complex functionality.

\tool supports \circom, \corset, \gnark, and \noir. It finds 13 bugs,
including seven with soundness impact. Several are reachable only
through generated standard-library calls. Under the same testing
budget, a valid-execution baseline does not expose any of the
soundness failures revealed by accepted injected witnesses, showing
that adversarial witness injection reaches failures missed by
valid-execution testing.
\end{abstract}

\input{intro}
\input{background}
\input{overview}
\input{lie-injection}
\input{circil}
\input{evaluation}
\input{related}
\input{conclusion}

%-------------------------------------------------------------------------------
\section*{Acknowledgments}
%-------------------------------------------------------------------------------

We are grateful to the ZK DSL toolchain developers for their valuable
help. This work was funded by the Austrian Science Fund (FWF) [Grant
  ID: 10.55776/DOC1345324] as well as by the Vienna Science and
Technology Fund (WWTF) and the City of Vienna [Grant ID:
  10.47379/ICT22007].

%-------------------------------------------------------------------------------
% optional clearing of the page
%% \cleardoublepage
%% \input{open}
%% \cleardoublepage

%% \input{ethics}
%% \cleardoublepage

\bibliographystyle{plain}
\bibliography{bibliography}

%\cleardoublepage

%\appendix

%%%%%%%%%%%%%%%%%%%%%%%%%%%%%%%%%%%%%%%%%%%%%%%%%%%%%%%%%%%%%%%%%%%%%%%%%%%%%%%%
\end{document}

%% file: intro.tex
%!TEX root = main.tex

%-------------------------------------------------------------------------------
\section{Introduction}
\label{sec:intro}
%-------------------------------------------------------------------------------

Zero-knowledge domain-specific languages (ZK DSLs) let developers
build verifiable applications without writing low-level constraints.
Their toolchains compile programs into constraint systems, generate
witnesses, and produce verifier-checked proofs. These systems combine
compilers, standard libraries, witness generators, and prover backends.
Bugs can cause \emph{soundness issues}, where invalid executions are
accepted, or \emph{completeness issues}, where valid ones are rejected.

Soundness bugs are especially severe. If generated constraints do not
enforce the source-program semantics, a malicious prover can prove a
nonexistent computation. In applications that use proofs for
authorization, this can enable unauthorized withdrawals or asset
minting, or bypass identity and policy checks. Completeness bugs can
instead deny service by preventing valid transactions or
time-sensitive proofs.

Existing ZK DSL testing techniques derive their oracles from valid,
semantics-preserving executions. \circuzz~\cite{HochrainerIsychev2025}
and MTZK~\cite{XiaoLiu2025} use metamorphic
relations~\cite{ChenCheung1998}, which should preserve observable
behavior. Such oracles can expose bugs affecting valid executions, but
do not directly test whether a toolchain rejects invalid witnesses.
Purely underconstrained computations may therefore remain invisible:
all valid executions behave correctly, although additional invalid
witnesses are accepted. Recent work finds that existing security tools
for zero-knowledge systems provide limited support for detecting such
semantic mismatches, in which constraints accept witnesses that do not
correspond to valid executions even when the computation is
deterministic~\cite{KolozyanSorger2026}. Chaliasos et
al.~\cite{ChaliasosAlFath2025} identify generating invalid witnesses
without violating unrelated constraints as an open challenge for
fuzzing ZK circuits. We address this challenge for ZK DSL toolchains.

\textbf{Our approach.}
We present \tool, a testing framework that generates ZK DSL programs
and tests toolchains with a new oracle: \emph{adversarial witness
injection}. For each deterministic program, \tool executes it on two
public input assignments with different outputs. It combines one
execution's input with the other's output and splices intermediate
values from both witnesses. The resulting witness is invalid by
construction, so a correct toolchain must reject it. Acceptance exposes
a soundness bug in enforcing the source-program semantics.

To reach realistic functionality, \tool generates parameterized calls
to standard-library routines. These routines often expand to many
constraints and intermediate witness values, where several bugs found
by \tool occur.

\begin{figure*}[t]
  \centering
  \resizebox{\textwidth}{!}{%
  \begin{tikzpicture}[
      font=\small\sffamily,
      node distance=0.9cm and 0.8cm,
      % Style for active processing stages
      stage/.style={
        draw=blue!80!black,
        fill=blue!5,
        thick,
        rounded corners=4pt,
        align=center,
        minimum width=1.8cm,
        minimum height=0.7cm
      },
      % Style for data artifacts / inputs / outputs
      data/.style={
        draw=teal!80!black,
        fill=teal!5,
        thick,
        align=center,
        minimum width=1.5cm,
        minimum height=0.7cm
      },
      % Arrow styles
      arr/.style={->, >=stealth, thick, draw=gray!80},
      lbl/.style={font=\scriptsize\sffamily, fill=white, inner sep=2pt}
    ]

    % --- Main Pipeline Nodes ---
    \node[data] (program) {Program\\$P$};
    \node[stage, right=of program] (compiler) {Compiler};
    \node[stage, right=of compiler] (witnessgen) {Witness\\Generator};
    \node[data, right=of witnessgen] (witness) {Witness\\$W$};
    \node[stage, right=of witness] (prover) {Prover};
    \node[data, right=of prover] (proof) {Proof\\$\pi$};
    \node[stage, right=of proof] (verifier) {Verifier};

    % --- Off-axis Nodes ---
    \node[data, below=1.0cm of witnessgen] (inputs) {Inputs\\$I$};
    \node[data, above=1.0cm of prover] (constraints) {Constraints\\$C$};
    
    % --- Verification Results (Moved to the right of Verifier) ---
    \node[draw=green!60!black, fill=green!10, thick, rounded corners=2pt, minimum width=0.6cm, minimum height=0.6cm, above right=0.1cm and 0.8cm of verifier.east] (valid) {\color{green!50!black}\textbf{\checkmark}};
    \node[draw=red!60!black, fill=red!10, thick, rounded corners=2pt, minimum width=0.6cm, minimum height=0.6cm, below right=0.1cm and 0.8cm of verifier.east] (invalid) {\color{red!50!black}\textbf{\scalebox{0.9}{$\times$}}};

    % --- Main Pipeline Edges ---
    \draw[arr] (program) -- (compiler);
    \draw[arr] (compiler) -- (witnessgen);
    \draw[arr] (witnessgen) -- (witness);
    \draw[arr] (witness) -- (prover);
    \draw[arr] (prover) -- (proof);
    \draw[arr] (proof) -- (verifier);
    
    % Split arrows going out to the results on the right
    \draw[arr] (verifier.east) -- ++(0.3,0) |- (valid.west);
    \draw[arr] (verifier.east) -- ++(0.3,0) |- (invalid.west);

    % --- Inputs Execution & Routing ---
    \draw[arr] (inputs) -- (witnessgen);
    
    % Public inputs now have the entire bottom area to themselves
    \draw[arr] (inputs.south) -- ++(0,-0.4) -| node[pos=0.25, lbl] {public} (verifier.south);

    % --- Constraints Routing ---
    \draw[arr] (compiler.north) |- (constraints.west);
    \draw[arr] (constraints) -- (prover);
    \draw[arr] (constraints.east) -| (verifier.north);

  \end{tikzpicture}%
  }
  \caption{Architecture of a ZK DSL toolchain. A program is compiled
    into constraints and executed on inputs to produce a witness. The
    prover uses both to produce a proof, which the verifier checks
    against the public inputs and constraints.}
  \label{fig:zk-toolchain}
\end{figure*}
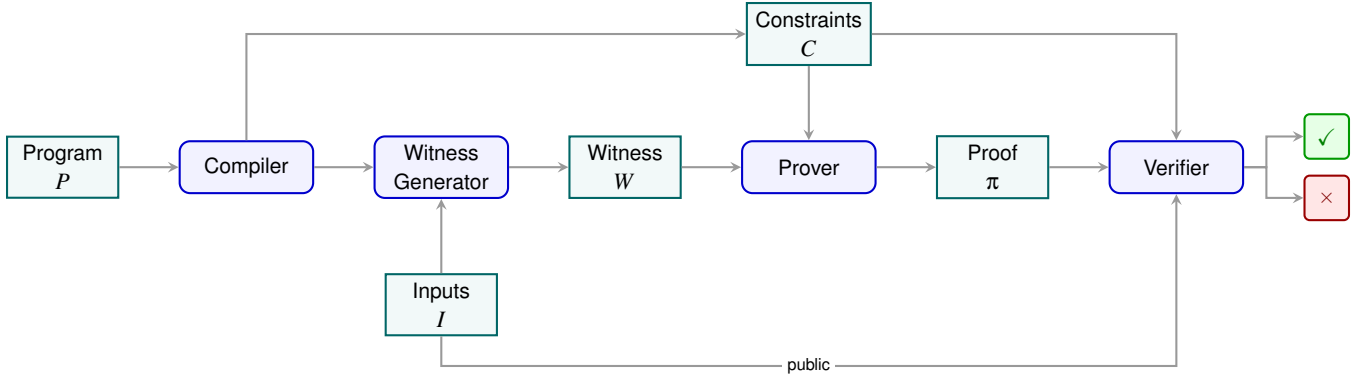

\textbf{Challenges.}
Randomly mutating a witness is insufficient. A mutation often violates
unrelated constraints, causing rejection even when part of the
computation is underconstrained. Moreover, the mutated witness is not
necessarily invalid because a constraint system may admit multiple
witnesses for the same source-level input-output behavior. For
example, changing an unused input produces a different but still valid
witness. \tool instead constructs invalid witnesses from valid
executions with different outputs and creates controlled divergence
between them. This preserves the expected witness structure while
introducing an input-output contradiction.

Complex operations may require different combinations of intermediate
values. \tool therefore uses multiple witness-splicing strategies to
expose both simple missing constraints and bugs requiring coordinated
changes to related values.

\textbf{Results.}
\tool supports four ZK DSL toolchains:
\circom~\cite{BellesMunozIsabel2023}, \corset~\cite{CorsetGo},
\gnark~\cite{Gnark}, and \noir~\cite{Noir}. It finds 13 bugs across
all toolchains, including seven with soundness impact. Several are
reachable only through generated standard-library calls. The \noir
soundness bug~\bug{9} in Table~\ref{tab:bugs} earned a \$70{,}000 bug
bounty.  Under the same budget, \circuzz's valid-execution oracle does
not expose the soundness failures revealed by accepted injected
witnesses, while remaining effective for completeness bugs. These
results show that adversarial witness injection targets a failure mode
that remains invisible when testing only valid executions.

\textbf{Contributions.}
We make the following contributions:
\begin{itemize}
\item We introduce adversarial witness injection, which constructs
  invalid witnesses from two valid executions using controlled
  divergence and multiple witness-splicing strategies.

\item We realize this oracle in \tool, supporting typed, parameterized
  standard-library calls and rich language constructs across four ZK
  DSL toolchains.

\item We evaluate \tool on four toolchains, showing that accepted
  injected witnesses expose soundness failures not observed by a
  valid-execution oracle under the same budget.
\end{itemize}

%%% Local Variables:
%%% mode: latex
%%% TeX-master: "main"
%%% End:

%% file: background.tex
%!TEX root = main.tex
%-------------------------------------------------------------------------------
\section{Background}
\label{sec:background}
%-------------------------------------------------------------------------------

%-------------------------------------------------------------------------------
\subsection{ZK DSL Toolchains}
\label{sec:background-toolchains}
%-------------------------------------------------------------------------------

Figure~\ref{fig:zk-toolchain} shows the architecture of a ZK DSL
toolchain. A program $P$ is compiled into a constraint system $C$.
Executing $P$ on inputs $I$ produces a witness $W$ containing values
that satisfy $C$. The prover uses $C$ and $W$ to produce a proof
$\pi$, which the verifier checks against the public inputs and $C$.

%-------------------------------------------------------------------------------
\subsection{Witnesses and Program Semantics}
\label{sec:background-witnesses}
%-------------------------------------------------------------------------------

For adversarial-witness testing, \tool generates programs with public
inputs and outputs. We write $P(I) \rightarrow O$ when executing $P$
on inputs $I$ produces outputs $O$. Abstractly, a witness contains
inputs, intermediate values $X$, and outputs:
\[
  W = [I \,\|\, X \,\|\, O]
\]
Witness layouts differ across toolchains, and some use execution traces
rather than flat arrays. \tool only needs to identify the inputs,
outputs, and intermediate values that it can modify.

\tool relies on determinism: fixed inputs determine outputs. If
$P(I_1) \rightarrow O_1$ and $P(I_2) \rightarrow O_2$ with
$O_1 \neq O_2$, a witness combining $I_1$ with $O_2$ cannot represent
a valid execution of $P$.

\begin{figure*}[t]
\centering

\begin{subfigure}[b]{0.44\textwidth}
\centering
\begin{lstlisting}[style=il]
inputs : i
outputs: o

o = fixed_base_scalar_mul_is_infinite(i)
      ? 123
      : 42
\end{lstlisting}
\caption{\il}
\end{subfigure}
\hfill
\begin{subfigure}[b]{0.54\textwidth}
\centering
\begin{lstlisting}[style=basic]
use std::embedded_curve_ops::{
    EmbeddedCurveScalar, fixed_base_scalar_mul};

pub fn main(i : pub(Field)) -> pub(Field) {
  let p = fixed_base_scalar_mul(
    EmbeddedCurveScalar::from_field(i)
  );
  let o = if (p.is_infinite) { 123 } else { 42 };
  (o)
}
\end{lstlisting}
\caption{\noir}
\end{subfigure}
\caption{Simplified version of a generated \il program for a \noir
  soundness bug found by \tool and its \noir translation. The program
  performs fixed-base scalar multiplication and branches on the
  resulting \code{is_infinite} flag.}
\label{fig:example}
\end{figure*}

%-------------------------------------------------------------------------------
\subsection{Soundness and Completeness Bugs}
\label{sec:background-bugs}
%-------------------------------------------------------------------------------

A completeness bug rejects a valid execution, for example through a
crash or rejection of a proof generated from a valid witness. A
soundness bug accepts an invalid execution, allowing a prover to
produce a proof that verifies even though the computation violates the
source-program semantics.

Adversarial witness injection targets soundness bugs by constructing
witnesses that a correct toolchain must reject. It can also expose
completeness bugs if the toolchain crashes while processing a generated
program or injected witness.

%-------------------------------------------------------------------------------
\subsection{\circuzz and \il}
\label{sec:background-circuzz}
%-------------------------------------------------------------------------------

\tool builds on the program-generation and translation infrastructure
of \circuzz, a metamorphic-testing framework for ZK DSL toolchains.
\circuzz generates valid, semantics-preserving executions and reports
crashes, rejected valid proofs, or inconsistent behavior. This oracle
detects bugs visible during valid executions, but does not test whether
invalid witnesses are rejected.

To support multiple ZK DSLs, \circuzz generates programs in the
intermediate language \il and translates them to each target language.
A \il program contains inputs, outputs, and typed operations over field
values and other DSL-level constructs. It is imperative and
first-order, with assignments and Boolean assertions. Arithmetic is
evaluated modulo the target field prime.

\tool reuses this common program representation and its
target-language translations. It changes the tested property from
consistency across valid executions to rejection of invalid witnesses,
and extends \il to generate the richer programs described in
Section~\ref{sec:il}.

%%% Local Variables:
%%% mode: latex
%%% TeX-master: "main"
%%% End:

%% file: overview.tex
%!TEX root = main.tex

%-------------------------------------------------------------------------------
\section{Overview}
\label{sec:overview}
%-------------------------------------------------------------------------------

This section gives a high-level view of adversarial witness injection
and the role of \il program generation. The core idea is to construct
witnesses that are invalid by construction yet retain much of the
structure of valid witnesses. This provides an oracle for whether a ZK
DSL toolchain rejects executions that violate the source-program
semantics.

\begin{figure*}[t]
  \centering
  \resizebox{0.95\textwidth}{!}{%
  \begin{tikzpicture}[
      font=\small\sffamily,
      node distance=0.6cm and 0.9cm,
      stage/.style={
        draw=blue!80!black,
        fill=blue!5,
        thick,
        rounded corners=4pt,
        align=center,
        minimum width=2.1cm,
        minimum height=0.8cm
      },
      data/.style={
        draw=teal!80!black,
        fill=teal!5,
        thick,
        align=center,
        minimum width=1.7cm,
        minimum height=0.8cm
      },
      witness/.style={
        draw=orange!80!black,
        fill=orange!8,
        thick,
        align=center,
        minimum width=2.7cm,
        minimum height=0.9cm
      },
      bug/.style={
        draw=red!70!black,
        fill=red!8,
        thick,
        rounded corners=4pt,
        align=center,
        minimum width=2.6cm,
        minimum height=0.8cm
      },
      ok/.style={
        draw=green!60!black,
        fill=green!8,
        thick,
        rounded corners=4pt,
        align=center,
        minimum width=2.6cm,
        minimum height=0.8cm
      },
      arr/.style={->, >=stealth, thick, draw=gray!80},
      lbl/.style={
        font=\scriptsize\sffamily,
        fill=white,
        inner sep=2pt,
        minimum height=0.4cm
      }
    ]

    % ==========================================
    % 1. CENTRAL EXECUTION BLOCK AND DATA TRACKS
    % ==========================================
    \node[
      stage,
      minimum height=3.2cm,
      minimum width=2.5cm
    ] (exec) {Compile and\\Execute Program};

    \node[data, left=1.6cm of exec, yshift=1.0cm]
      (i1) {Inputs\\$I_1$};
    \node[data, left=1.6cm of exec, yshift=-1.0cm]
      (i2) {Inputs\\$I_2$};

    % ==========================================
    % 2. PROGRAM GENERATION AND TRANSLATION
    % ==========================================
    \node[stage, above=1.6cm of i1, xshift=0.2cm]
      (generate) {Generate\\\il Program};
    \node[data, right=of generate]
      (ilprog) {\il Program\\$P$};
    \node[stage, right=of ilprog]
      (translate) {Translate to\\Target DSL};
    \node[data, right=of translate]
      (target) {Target\\Program};
    \node[stage, right=of target]
      (choose) {Sample\\Inputs};

    % ==========================================
    % 3. SPLICING AND WITNESSES
    % ==========================================
    \node[
      stage,
      minimum width=2.2cm,
      minimum height=2.8cm,
      right=3.8cm of exec
    ] (splice) {Splice\\Witnesses};

    \coordinate (splicein1) at ([yshift=0.75cm]splice.west);
    \coordinate (splicein2) at ([yshift=-0.75cm]splice.west);

    \node[witness, left=0.6cm of splicein1, anchor=east]
      (w1) {$W_1$\\$[I_1 \,\|\, X_1 \,\|\, O_1]$};
    \node[witness, left=0.6cm of splicein2, anchor=east]
      (w2) {$W_2$\\$[I_2 \,\|\, X_2 \,\|\, O_2]$};

    \node[lbl, minimum width=1.4cm]
      (condition) at ($(w1.south)!0.5!(w2.north)$)
      {$O_1 \neq O_2$};

    % ==========================================
    % 4. LYING WITNESS AND CHECKING
    % ==========================================
    \node[witness, right=of splice]
      (wlie) {$W_{\mathit{lie}}$\\
      $[I_1 \,\|\, \mathit{mix}(X_1,X_2) \,\|\, O_2]$};

    \node[stage, right=of wlie]
      (check) {Check Injected\\Witness};

    \node[bug, right=0.7cm of check]
      (accept) {Accept\\Soundness bug};
    \node[ok, above=0.4cm of accept]
      (reject) {Reject\\Expected};
    \node[bug, below=0.4cm of accept]
      (crash) {Crash\\Completeness bug};

    % ==========================================
    % ROUTING AND CONNECTIONS
    % ==========================================

    \draw[arr] (generate) -- (ilprog);
    \draw[arr] (ilprog) -- (translate);
    \draw[arr] (translate) -- (target);
    \draw[arr] (target) -- (choose);

    \draw[arr]
      (choose.south) -- ++(0,-0.7)
      -| ([xshift=-0.4cm]i2.west) |- (i2.west);
    \draw[arr]
      (choose.south) -- ++(0,-0.7)
      -| ([xshift=-0.4cm]i1.west) |- (i1.west);

    \draw[arr]
      (target.south) -- ++(0,-0.2) -| (exec.north);

    \draw[arr]
      (i1.east) --
      node[lbl, pos=0.5] {$P(I_1)\!\to\!O_1$}
      ([yshift=1.0cm]exec.west);
    \draw[arr]
      (i2.east) --
      node[lbl, pos=0.5] {$P(I_2)\!\to\!O_2$}
      ([yshift=-1.0cm]exec.west);

    \draw[arr]
      ([yshift=0.75cm]exec.east) -- (w1.west);
    \draw[arr]
      ([yshift=-0.75cm]exec.east) -- (w2.west);

    \draw[arr] (w1.east) -- (splicein1);
    \draw[arr] (w2.east) -- (splicein2);
    \draw[arr] (splice) -- (wlie);
    \draw[arr] (wlie) -- (check);

    \draw[arr]
      (check.east) -- ++(0.2,0) |- (reject.west);
    \draw[arr]
      (check.east) -- (accept.west);
    \draw[arr]
      (check.east) -- ++(0.2,0) |- (crash.west);

    % Compilation or witness-generation failures.
    \draw[arr]
      (exec.south) -- ++(0,-0.55) -| (crash.south);

  \end{tikzpicture}%
  }
  \caption{\tool workflow for adversarial witness injection. \tool
    generates and translates a test program, compiles and executes it
    on pairs of inputs, and retains pairs with different outputs. It
    splices their witnesses to construct a lying witness. Rejection is
    expected; acceptance exposes a soundness bug, while a crash during
    compilation, witness generation, or processing of the injected
    witness exposes a completeness bug.}
  \label{fig:tool-workflow}
\end{figure*}
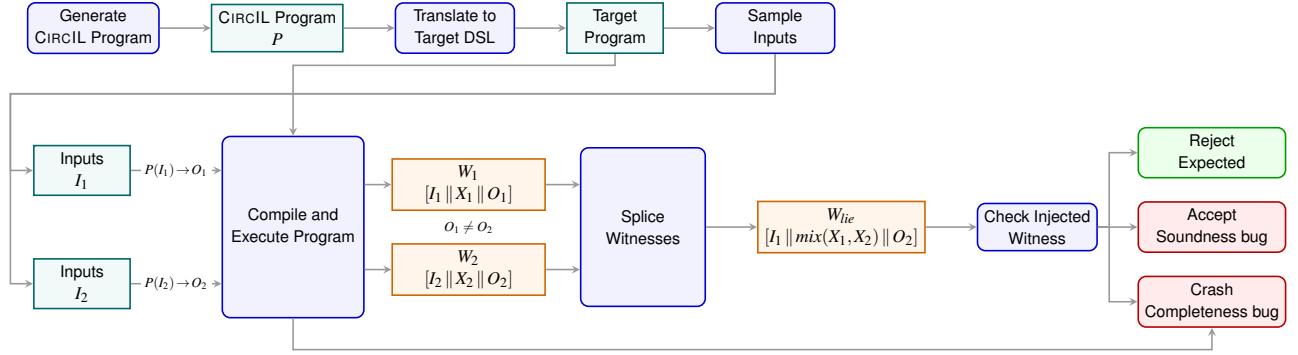

%-------------------------------------------------------------------------------
\subsection{Running Example}
\label{sec:overview-example}
%-------------------------------------------------------------------------------

We illustrate adversarial witness injection with a real \noir bug
found by \tool. Figure~\ref{fig:example} shows a simplified version of
a generated \il program and its \noir translation. The program
performs fixed-base scalar multiplication and branches on whether the
resulting point is the point at infinity. The \noir translation calls
the corresponding standard-library routine and uses its
\code{is\_infinite} flag. Function support in \il is essential here:
without it, \tool could not generate programs that reach this routine.

For scalar $0$, fixed-base scalar multiplication yields the point at
infinity, so \code{is\_infinite} should be true and the program should
return $123$. For scalar $1$, it yields the curve generator, so the
flag should be false and the program should return $42$. The bug is
that \noir constrains the point coordinates but not their relationship
with \code{is\_infinite}, allowing the flag to be changed
independently. The toolchain can therefore accept invalid witnesses,
making this a soundness bug. We reported the bug, and the developers
fixed it.

Consider two executions of the program. We write the relevant
intermediate values as $p_x$, $p_y$, and $b$, where $p_x$ and $p_y$ are
the point coordinates and $b$ encodes \code{is\_infinite}. On input
$i=1$, the routine returns the generator $(g_x,g_y)$ with $b=0$, and
the program returns $42$:
\[
  W_1 =
  [1 \,\|\, \ldots, p_x=g_x, p_y=g_y, b=0 \,\|\, 42]
\]
On input $i=0$, it returns the point at infinity, represented by
coordinates $(0,0)$ with $b=1$, and the program returns $123$:
\[
  W_2 =
  [0 \,\|\, \ldots, p_x=0, p_y=0, b=1 \,\|\, 123]
\]
Both witnesses represent valid executions.

We call an adversarially constructed invalid witness a
\emph{lying witness}. \tool constructs one by retaining the input and
point coordinates from the first execution while taking the flag and
output from the second:
\[
  W_{\mathit{lie}} =
  [1 \,\|\, \ldots, p_x=g_x, p_y=g_y, b=1 \,\|\, 123]
\]
The lying witness is invalid. It claims that input $1$ produces the
generator $(g_x,g_y)$ but also that this point is the point at
infinity. No valid execution can satisfy both claims. A correct
constraint system must therefore reject the witness. The buggy system
accepts it because it constrains the coordinates but not their
relationship with \code{is\_infinite}.

The example illustrates two ingredients of \tool. First, function
support in \il generates programs that reach standard-library code.
Second, adversarial witness injection exercises behavior absent from
valid executions. Both $W_1$ and $W_2$ are valid, and the program
behaves correctly on them. Only their adversarial combination exposes
the bug. Sections~\ref{sec:lie-injection} and~\ref{sec:il} describe
these ingredients in detail.

%-------------------------------------------------------------------------------
\subsection{\tool Workflow}
\label{sec:overview-workflow}
%-------------------------------------------------------------------------------

Figure~\ref{fig:tool-workflow} shows how \tool combines \il program
generation, translation, input sampling, and adversarial witness
injection. \tool generates a \il program and translates it to a
target ZK DSL. It then samples pairs of input assignments, compiles
the translated program, and executes it on each assignment to obtain
two witnesses. It retains only pairs whose outputs differ. Such pairs
let \tool combine an input with an output that cannot belong to the
same execution of the deterministic
program. Section~\ref{sec:lie-divergence} explains how \tool
constructs these input pairs.

For each retained pair, \tool selects a witness-splicing strategy and
constructs a lying witness. It then checks the witness with the target
toolchain. Rejection is expected. Acceptance means that the toolchain
accepted an execution that violates the source-program semantics,
exposing a soundness bug. A crash during compilation, witness
generation, or processing of the injected witness exposes a
completeness bug.

%%% Local Variables:
%%% mode: latex
%%% TeX-master: "main"
%%% End:

%% file: lie-injection.tex
%!TEX root = main.tex

%-------------------------------------------------------------------------------
\section{Adversarial Witness Injection}
\label{sec:lie-injection}
%-------------------------------------------------------------------------------

This section presents adversarial witness injection in \tool. The
oracle relies on deterministic source semantics: for a fixed input
assignment, executing the generated program determines its output. For
adversarial-witness testing, \tool generates programs with public inputs
and outputs. It then tests whether the target toolchain correctly
enforces this input-output relation.

%-------------------------------------------------------------------------------
\subsection{Constructing Lying Witnesses}
\label{sec:lie-witnesses}
%-------------------------------------------------------------------------------

The oracle follows directly from determinism. Suppose \tool executes a
program $P$ twice and obtains
\[
  P(I_1) \to O_1
  \qquad\text{and}\qquad
  P(I_2) \to O_2 ,
\]
with $O_1 \neq O_2$. Since $P$ is deterministic, $O_2$ cannot be the
output of $P$ on input $I_1$. Any witness combining $I_1$ with $O_2$ is
therefore invalid, regardless of its intermediate values.

Let the two valid witnesses be
\[
  W_1 = [I_1 \,\|\, X_1 \,\|\, O_1]
  \qquad\text{and}\qquad
  W_2 = [I_2 \,\|\, X_2 \,\|\, O_2],
\]
using the abstract witness notation from
Section~\ref{sec:background-witnesses}. \tool constructs the lying
witness
\[
  W_{\mathit{lie}} =
  [I_1 \,\|\, \mathit{mix}(X_1,X_2) \,\|\, O_2] .
  \]
The input of $W_{\mathit{lie}}$ comes from the first execution, while
its output comes from the second. The function $\mathit{mix}$ determines
how intermediate values are selected from the two valid witnesses. We
call the overall construction \emph{witness splicing} and the choice
encoded by $\mathit{mix}$ the \emph{splicing strategy}. Regardless of
this choice, the combination of $I_1$ and $O_2$ makes
$W_{\mathit{lie}}$ invalid.

A correct toolchain must reject $W_{\mathit{lie}}$. We use
\emph{acceptance} for the target-specific outcome in which the injected
witness passes the toolchain's check. Acceptance means that the
toolchain has accepted an execution that cannot occur under the
source-program semantics, so \tool reports a soundness bug. If the
toolchain crashes during compilation, while generating the original
witnesses, or while processing the injected witness, \tool reports a
completeness bug.

%-------------------------------------------------------------------------------
\subsection{Generating Diverging Executions}
\label{sec:lie-divergence}
%-------------------------------------------------------------------------------

Before splicing witnesses, \tool must obtain two valid executions whose
outputs differ. A simple approach is to sample two unrelated inputs
$I_1$ and $I_2$ and retain the pair if $P(I_1) \neq P(I_2)$. This is
sufficient for the oracle, but often ineffective at exposing missing
constraints.

Unrelated inputs can cause many computations to differ simultaneously.
Even changing a single input can have this effect if the program uses
it in several places. The resulting lying witness may then violate
constraints unrelated to the missing constraint that \tool is trying
to expose. For example, suppose the same input is used in both a
multiplication and an addition. Even if the multiplication constraint
is missing, the addition constraint may reject a spliced witness
because it still enforces consistent use of the changed input. The
addition therefore masks the missing multiplication constraint.

\tool instead tries to create a single controlled divergence point.
During \il program generation, it selects one occurrence of an existing
input and replaces that occurrence with a fresh \emph{fault input} $f$.
It then constructs two input assignments that agree on every ordinary
input and differ only in the value assigned to $f$:
\[
  I_1(x) = I_2(x)
  \quad\text{for every input } x \neq f,
  \qquad
  I_1(f) \neq I_2(f)
\]
The two executions follow the same computation until reaching the
selected occurrence of $f$, at which point their intermediate values
may diverge. This controlled divergence localizes the differences
between the witnesses. Constraints that do not depend on the
occurrence are more likely to remain satisfied, while the lying
witness still combines an input with an output that cannot belong to
the same deterministic execution.

\tool chooses the second value of $f$ by applying several mutation
strategies. It tries local changes such as $f_2 = f_1 + 1$ and $f_2 =
f_1 - 1$, sign changes such as $f_2 = -f_1$, special field values
$0$, $1$, and $p-1$, where $p$ is the field modulus, and random
offsets. After executing the program on both assignments, \tool retains
the pair only if the public outputs differ. Pairs with equal outputs are
discarded because they do not provide the required input-output
contradiction.

%-------------------------------------------------------------------------------
\subsection{Witness-Splicing Strategies}
\label{sec:lie-splicing}
%-------------------------------------------------------------------------------

The construction in Section~\ref{sec:lie-witnesses} fixes the inputs
and outputs of the lying witness but leaves open how to choose its
intermediate values. A missing constraint is most likely to be exposed
when the lying witness preserves enough consistency to satisfy
unaffected constraints while still combining values that cannot occur
together in a valid execution.

The difficulty is that \tool does not know which frontend witness
values are constrained together. Standard-library functions may
produce several witness values, and a missing constraint may be
exposed only by particular combinations from the two executions. The
running example in Section~\ref{sec:overview-example} illustrates this
challenge: exposing the \noir bug requires retaining the point
coordinates from one execution while taking the \code{is\_infinite}
flag from the other.

\tool therefore supports several ways of instantiating
$\mathit{mix}(X_1,X_2)$. For each retained pair of executions, it
selects a splicing strategy, constructs a lying witness, and checks the
witness with the target toolchain. Each strategy relies on controlled
divergence: the two executions agree until the selected use of the fault
input $f$ and may differ afterwards. We call the first intermediate
position at which their witness values may differ the
\emph{divergence point}. Values before this point are copied from
$X_1$, because they are common to both executions. The strategies
differ in how they choose values at and after the divergence point.

\textbf{Divergence-point splicing.}
This strategy switches from $X_1$ to $X_2$ exactly at the divergence
point. The lying witness retains the common prefix from the first
execution and takes all remaining intermediate values from the second.
This is the most direct instantiation of $\mathit{mix}(X_1,X_2)$.

\textbf{Shifted splicing.}
This strategy delays the switch from $X_1$ to $X_2$ by a small offset.
The lying witness retains the common prefix and a bounded number of
post-divergence values from $X_1$, then takes the remaining
intermediate values from $X_2$. The offset is chosen from a bounded
range.

\textbf{Windowed splicing.}
This strategy defines a bounded window starting at the divergence
point. Outside the window, the lying witness follows divergence-point
splicing. Inside it, \tool mixes values from $X_1$ and $X_2$. In the
uniform variant, each position is selected from either execution with
a fixed probability. In the tapered variant, the probability of
selecting values from $X_1$ decreases across the window, creating a
gradual transition from the first execution to the second.

\textbf{Suffix splicing.}
This strategy extends the mixing region from the divergence point to
the end of the intermediate-value sequence. As with windowed splicing,
\tool uses uniform and tapered variants. Suffix splicing can therefore
retain values from $X_1$ far beyond the divergence point rather than
confining the mixture to a bounded window.

Together, these strategies explore abrupt, delayed, localized, and
gradual transitions between the two executions. For each test, \tool
checks the lying witness constructed by the selected strategy and
reports a soundness bug if the target toolchain accepts it.

%%% Local Variables:
%%% mode: latex
%%% TeX-master: "main"
%%% End:

%% file: circil.tex
%!TEX root = main.tex

%-------------------------------------------------------------------------------
\section{Extending \il}
\label{sec:il}
%-------------------------------------------------------------------------------

Adversarial witness injection is most effective when generated programs
exercise functionality beyond primitive arithmetic. Many bugs in ZK DSL
toolchains occur in standard-library routines, conversions,
parameterized operations, and other constructs that expand into many
constraints and intermediate witness values. To reach such code, \tool
extends \il with typed function specifications, generation-time
constraints, and supporting language constructs.

%-------------------------------------------------------------------------------
\subsection{Function Specifications}
\label{sec:il-functions}
%-------------------------------------------------------------------------------

The most important \il extension for \tool is support for function
specifications. In \circuzz, generated programs were built primarily
from primitive \il operations. This was sufficient for testing many
toolchain behaviors, but limited the generated programs to functionality
expressible directly in the base language. Many ZK DSLs, however,
provide important functionality through standard-library routines,
including curve operations, conversions, comparisons, decompositions,
and hash functions. Bugs in these routines cannot be reached unless
generated programs can call them.
Such calls exercise both the source-level library implementations and
their translation through later toolchain stages.

\tool therefore extends \il with typed function specifications. A
function specification describes the arguments and return values of a
routine that may be called from a generated \il program. A
target-specific translation maps the \il call to the corresponding
standard-library routine or language construct in the target ZK DSL.
This lets \tool exercise functionality beyond the primitive \il
operations while keeping program generation independent of
target-language syntax.

Function specifications can also be parameterized. For example, an
array-concatenation specification can be parameterized by an element
type $T$ and three length parameters $N$, $M$, and $S$:
\[
  \mathsf{concat}\langle T,N,M,S\rangle :
    [T;N] \times [T;M] \to [T;S]
\]
Here, $T$ is a generic type, $N$ and $M$ are the input-array lengths,
and $S$ is the output-array length. Their relationship is enforced by
generation-time constraints, discussed in
Section~\ref{sec:il-constraints}. When \tool instantiates the
specification, it chooses a concrete element type and concrete lengths
that satisfy these constraints, then emits the corresponding \il call.
Other specifications use parameters to select properties such as array
lengths, integer widths, signedness, and element types.

Function support is central to several bugs found by \tool. The
example in Section~\ref{sec:overview-example} uses a generated call to
a fixed-base scalar-multiplication routine. Without function
specifications, \tool could not generate the \il program that reaches
this standard-library code. The same mechanism lets \tool exercise
other standard-library functionality, including conversions, hash
functions, and scalar-field operations.

%-------------------------------------------------------------------------------
\subsection{Generation-Time Constraints}
\label{sec:il-constraints}
%-------------------------------------------------------------------------------

Parameterized function specifications introduce a second challenge:
not every choice of generic parameters describes a valid
target-language call. Array sizes may need to be nonzero, integer
widths may be restricted by the target toolchain, and relationships
between parameters may need to hold. Without checking these conditions,
many generated programs would fail for uninteresting type or
configuration reasons before reaching the behavior under test.

\tool therefore attaches generation-time constraints to function
specifications. These constraints are evaluated during \il program
generation, before a program is emitted. They restrict the generic
values chosen when a parameterized specification is instantiated. They
may impose constant bounds, such as requiring an array length to be at
least one, or express relationships between multiple generic values.

For example, the array-concatenation specification from
Section~\ref{sec:il-functions} has three length parameters: the input
lengths $N$ and $M$, and the output length $S$. A generated call is
valid only when these parameters describe a consistent concatenation:
\[
\begin{array}{l}
  \mathsf{concat}\langle T,N,M,S\rangle :
    [T;N] \times [T;M] \to [T;S] \\[2pt]
  0 \leq N \leq S, \qquad M = S - N
\end{array}
\]
These constraints are checked during \il program generation, so \tool
emits only calls whose lengths satisfy target requirements. They are
not source-level assertions and do not become part of the generated
ZK DSL program. Instead, they guide generation and avoid invalid
instantiations that would fail during type checking or configuration
validation.

%-------------------------------------------------------------------------------
\subsection{Additional Program Constructs}
\label{sec:il-constructs}
%-------------------------------------------------------------------------------

Function specifications require \il to represent the values consumed
and produced by standard-library routines. In addition to field and
Boolean values, \tool therefore extends \il with bounded-size arrays.
Arrays may appear as program inputs and outputs, as function arguments,
and as return values. The generator can construct array literals,
assign whole arrays, read or update individual elements, and initialize
arrays element by element.

Arrays are especially useful for functions whose natural interface is
not a single field element. For example, list operations, hash
functions, and field-to-bit or field-to-byte conversions operate on
collections of values. Combined with generic function specifications,
arrays let one specification range over both the element type and the
array length, as in the \codeil{concat} example from
Section~\ref{sec:il-functions}.

\tool also adds let expressions, which introduce expression-local
bindings. Unlike statement-level assignments, a let expression
evaluates to a value and can therefore appear within a larger
expression. This lets the generator create additional intermediate
values and more structured expressions. For \noir, the translation
randomly chooses between introducing an intermediate variable and
encoding the let expression as an inline closure. The latter exercises
target-language machinery not reached by ordinary statement-level
assignments.

%%% Local Variables:
%%% mode: latex
%%% TeX-master: "main"
%%% End:

%% file: evaluation.tex
%!TEX root = main.tex

%-------------------------------------------------------------------------------
\section{Evaluation}
\label{sec:evaluation}
%-------------------------------------------------------------------------------

We evaluate \tool on four ZK DSL toolchains: \circom, \corset, \gnark,
and \noir. Our evaluation addresses five research questions:

\begin{description}
  \item[RQ1:] How effective is \tool at finding soundness and
    completeness bugs?
  \item[RQ2:] What kinds of bugs does adversarial witness injection
    expose?
  \item[RQ3:] How reliably and quickly does \tool rediscover the
    reported bugs?
  \item[RQ4:] What does adversarial witness injection expose beyond
    \circuzz's metamorphic-testing oracle?
  \item[RQ5:] How do the new \il program-generation features affect
    bug-finding effectiveness?  %
\end{description}

%-------------------------------------------------------------------------------
\subsection{Experimental Setup}
\label{sec:eval-setup}
%-------------------------------------------------------------------------------

\textbf{Target toolchains.}
We evaluate \tool on \circom, \corset, \gnark, and \noir. These
toolchains differ in their source languages and workflows.
For each toolchain, \tool excludes source constructs that intentionally
leave values unconstrained, such as \circom's unconstrained assignments.
For \circom, we additionally exclude operations that cannot be expressed
automatically as quadratic constraints.

Our \corset integration targets the zkASM language, whereas \circuzz
targeted \corset's Lisp-like constraint language~\cite{Corset}. A
zkASM program is compiled into a constraint system and executed to
produce an execution trace that serves as the witness. Unlike the
other evaluated toolchains, \corset does not produce a cryptographic
proof in our workflow. Instead, we use its checker to determine
whether a generated or injected trace satisfies the compiled
constraints.

In adversarial-witness campaigns, generated program
inputs and outputs are public for \circom, \gnark, and \noir; zkASM
does not distinguish public from private values.

\textbf{Witness representations.}
\circom and \noir expose witnesses as flat arrays whose entries can be
mapped to inputs, outputs, and intermediate values. \gnark also uses a
flat witness array, but does not expose the full array through its
public interface. We therefore modify \gnark to extract the
intermediate witness values and to supply an injected witness during
proving.

\corset instead represents its witness as a two-dimensional trace. We
adapt the witness-splicing strategies from
Section~\ref{sec:lie-splicing} to combine two valid traces. Because
\corset can check candidate traces quickly without generating a proof,
we additionally use a brute-force strategy that constructs many
candidate traces from values in the two valid executions. Each
candidate preserves the required input-output contradiction and is
submitted to the checker to determine whether the compiled constraints
accept it.

\textbf{Generated programs.}
We generate relatively small \il programs to maximize testing
throughput. The default configuration uses one or two inputs, up to
two outputs, and expressions with a maximum depth of four. In
adversarial-witness campaigns, we disable explicit \il assertions
because a failed assertion would prevent \tool from obtaining the two
valid executions needed to construct a lying witness.

\textbf{Toolchain versions.}
We used the latest main-branch version of each toolchain available at
the time of testing, together with release or main-branch versions of
separate prover components.

\textbf{Hardware.}
The experiments were run on a machine with an AMD EPYC 9474F processor
and 1.5\,TB of memory, running Debian GNU/Linux 13 (trixie).

%-------------------------------------------------------------------------------
\subsection{RQ1: Bugs Found}
\label{sec:eval-bugs}
%-------------------------------------------------------------------------------

\begin{table*}[t]
  \centering
  \caption{Bugs found by \tool. Linked bug numbers point to public
    reports or advisories, and linked fix letters point to public pull
    requests or commits.}
  \label{tab:bugs}
  \begin{tabularx}{\textwidth}{cclllX}
    \toprule
    \textbf{Bug} & \textbf{Fix} & \textbf{Toolchain} &
    \textbf{Type} & \textbf{Signal} & \textbf{Description} \\
    \midrule
    \bug{1}  & \fix{A} & \circom & both         & accepted lie  & Sign flip in R1CS-to-Plonk constraint translation. \\
    \midrule
    \bug{2}  & --      & \corset & soundness    & accepted lie  & Incorrect culling of complex constraints. \\
    \bug{3}  & \fix{B} & \corset & soundness    & accepted lie  & Missing constraints after return statements. \\
    \bug{4}  & \fix{C} & \corset & completeness & check crash   & Checker error during limb-width calculation. \\
    \bug{5}  & --      & \corset & completeness & compile crash & Compiler crash on small fields. \\
    \bug{6}  & --      & \corset & completeness & check rejection & Valid traces rejected by generated constraints. \\
    \midrule
    \bug{7}  & \fix{D} & \gnark  & soundness    & accepted lie  & Unconstrained hints in curve scalar multiplication. \\
    \bug{8}  & \fix{E} & \gnark  & completeness & compile crash & Crash on effectively constant lookup entry. \\
    \midrule
    \bug{9}  & \fix{F} & \noir   & soundness    & accepted lie  & Incorrect lowering of large ACIR constraints. \\
    \bug{10} & \fix{G} & \noir   & soundness    & accepted lie  & Unconstrained \code{is\_infinite} flag in curve points. \\
    \bug{11} & \fix{H} & \noir   & completeness & compile crash & Crash when exporting verification key. \\
    \bug{12} & \fix{I} & \noir   & completeness & compile crash & Assertion error during scalar decomposition. \\
    \bug{13} & \fix{J} & \noir   & both         & witness crash & Incorrect removal of cast statements. \\
    \bottomrule
  \end{tabularx}
\end{table*}

We conducted open-ended bug-finding campaigns as we added support for
each toolchain. If a detected bug prevented further testing, we paused
the campaign until the bug was fixed or worked around. These
exploratory campaigns establish the set of discovered bugs but are not
used for quantitative comparisons between toolchains.
Table~\ref{tab:bugs} summarizes the bugs found by \tool.
The \emph{Bug} column identifies each bug and links to the
corresponding public report or advisory when available; an unlinked
number denotes a bug without a public report. The \emph{Fix} column
identifies the fix. Linked letters point to public pull requests or
commits. A dash indicates that no fix is available.

The \emph{Type} column records the impact of the bug: soundness,
completeness, or both. The \emph{Signal} column records the
observation that exposed the bug. For soundness bugs, the signal is
that the toolchain accepted an injected witness that should have been
rejected. For completeness bugs, possible signals include a crash during
compilation, while generating the original witnesses, or while
processing the injected witness. For \corset, processing the injected
witness means checking the trace against the generated constraints;
rejection of a valid trace is also a completeness signal.

Overall, \tool found 13 bugs across the four evaluated toolchains:
five soundness bugs, six completeness bugs, and two bugs with both
soundness and completeness impact. Six of the seven bugs with
soundness impact were exposed when the toolchain accepted a lying
witness. Bug~\bug{13} was first observed as a crash but can also cause
a soundness failure. The bugs affect several major parts of the tested
toolchains, including compilers, witness generators, provers, and
constraint-system checkers for \corset.

The soundness bugs are the most security-critical results. In
proof-producing workflows, they allow a malicious prover to
produce a proof that verifies for a witness that does not match the
source-program semantics. Such bugs can directly undermine
applications that rely on ZK proofs for authorization, asset
transfers, membership checks, or other security-sensitive
decisions. Completeness bugs are less directly exploitable, but still
matter in deployed systems: they can prevent users from compiling
programs, generating witnesses, or producing proofs that verify, and
can therefore cause denial of service in protocols that depend on
timely proof generation.

All bugs with soundness impact in Table~\ref{tab:bugs} except
bug~\bug{2} were fixed by the corresponding developers. The remaining
issue is a confirmed \corset MIR bug. The developers indicated that
this issue is not a priority because the affected representation is
not used in production.

%-------------------------------------------------------------------------------
\subsection{RQ2: Bug Characteristics}
\label{sec:eval-characteristics}
%-------------------------------------------------------------------------------

This section characterizes the bugs found by \tool. We focus primarily
on the bugs with soundness impact, because they are the most
security-critical results and the main target of adversarial witness
injection. We discuss completeness bugs where they illustrate a distinct
failure mode or show the importance of the new \il program-generation
features.

\textbf{Underconstrainedness.}
The main pattern is that the soundness bugs found through accepted
injected witnesses are underconstrainedness bugs: the generated
constraints accept more witnesses than the source-program semantics
permits. In such cases, valid executions can still behave correctly,
because the toolchain accepts all valid witnesses. The bug becomes
visible only when \tool supplies an invalid witness that preserves
enough structure to satisfy the existing constraints while violating the
source-level input-output relation.

The soundness bugs occur in different parts of the tested toolchains.
In \corset, bugs~\bug{2} and~\bug{3} produce constraint systems that
are too weak: one incorrectly culls complex constraints, while the
other omits constraints needed after return statements. In \gnark,
bug~\bug{7} affects a standard-library curve operation whose
hint-based scalar decomposition can be satisfied by malicious witness
values. In \noir, bug~\bug{9} weakens constraints when Barretenberg
lowers ACIR, \noir's intermediate representation, into backend
constraints. Bug~\bug{10}, the running example from
Section~\ref{sec:overview-example}, leaves the \code{is\_infinite}
flag unconstrained with respect to the point coordinates.

Most bugs exposed by accepted injected witnesses are missing-constraint
rather than wrong-constraint bugs. The fixes therefore add constraints
or prevent invalid values from being used, rather than changing the
semantics of valid executions. This explains why semantics-preserving
testing, as in \circuzz, is not expected to expose them: if valid
witnesses are accepted and produce the expected outputs, an oracle based
only on valid executions has no disagreement to observe. Bug~\bug{1} is
the main exception. It is caused by a sign error when the \circom prover
converts R1CS constraints into its internal Plonk representation, so it
can both accept invalid witnesses and reject valid ones.

\textbf{Witness-splicing behavior.}
The bugs show why \tool uses controlled divergence and multiple
witness-splicing strategies. The successful strategy depends on how
much post-divergence consistency the buggy constraint system requires,
and on whether the fault-input mutation reaches an edge case. Some bugs
are exposed by switching from one execution to the other.
Bug~\bug{7} is exposed by divergence-point splicing, typically with a
mutation that makes one execution use scalar zero. This matches the
structure of the bug, because the curve operation has an exceptional
scalar-zero case.

In contrast, bug~\bug{10} requires a more selective combination of
intermediate values. \tool must keep the point coordinates from one
execution while taking the \code{is\_infinite} flag and output from
the other. Accordingly, successful tests use shifted, windowed, or
suffix splicing rather than divergence-point splicing, together with
fault-input mutations to special field values such as zero. These
strategies can keep post-divergence values from the first execution
while changing others, which is necessary when the missing constraint
relates only part of a larger structured value.

Bug~\bug{9} shows strategy-insensitive underconstrainedness. All
witness-splicing strategies and many fault-input mutations expose it.
This reflects the bug's severity: the toolchain accepts many
combinations of witness values that do not correspond to valid
source-level executions.

Two further bugs were rediscovered only rarely. Bug~\bug{1} was
rediscovered only in the ablated \tool configuration evaluated in RQ5.
The successful test used the sign-changing mutation $f_2=-f_1$ and
divergence-point splicing. Bug~\bug{3} was rediscovered once with the
full configuration (RQ3) and once with the ablated configuration
(RQ5), both using the brute-force trace strategy.  The successful
tests used $f_2=f_1-1$ and $f_2=0$, respectively.

\textbf{Standard-library reachability.} Standard-library functionality
is a source of security bugs. Bugs~\bug{7} and~\bug{10} both affect
elliptic-curve operations and require programs to call library
routines. They would not be reachable with the base \il program
generation, which did not emit calls to standard-library routines. The
crash bugs show a related pattern for completeness: bug~\bug{8}
involves lookup tables, while bug~\bug{11} is triggered by redundant
target-level assertions that can arise from richer expressions and
standard-library calls, and bug~\bug{12} requires elliptic-curve
scalar operations. These bugs motivate the \il extensions from
Section~\ref{sec:il}: without function specifications and richer
programs, \tool would not exercise the library routines and expression
patterns that trigger several of the bugs in Table~\ref{tab:bugs}.

\textbf{Mixed soundness and completeness impact.}
Finally, the table includes two bugs with both soundness and
completeness impact. Bug~\bug{1} accepts witnesses satisfying an
incorrect sign-flipped constraint and rejects witnesses satisfying the
intended constraint. Bug~\bug{13} was first observed through a crash,
but the underlying cast-removal bug can also produce incorrect
constraints. These cases show that the boundary between soundness and
completeness is not always tied to a single symptom: the same root cause
can both reject valid executions and accept invalid ones, depending on
how it is triggered.

%-------------------------------------------------------------------------------
\subsection{RQ3: Time to Bug}
\label{sec:eval-ttb}
%-------------------------------------------------------------------------------

\begin{table*}[t]
  \centering
  \caption{Time to bug for fixed bugs. The \emph{Seeds} column reports
    the number of successful campaigns out of ten. \emph{Time} and
    \emph{\slantbox{\il} programs} report minimum/median/maximum values among
    successful campaigns; sub-minute times are shown as \($<$\)1m. Bugs above the horizontal line were
    originally exposed through accepted injected witnesses; bugs below
    it were originally exposed through crashes or rejections.}
  \label{tab:ttb}
  \begin{tabularx}{\textwidth}{X c l r X r r r X r r r X }
    \toprule
    & \multirow{2.5}{*}{\textbf{Bug}} & \multirow{2.5}{*}{\textbf{Toolchain}} & \multirow{2.5}{*}{\textbf{Seeds}} & & \multicolumn{3}{c}{\textbf{Time to bug}} & & \multicolumn{3}{c}{\textbf{\il programs}} & \\
    \cmidrule(lr){6-8} \cmidrule(lr){10-12} & & & & & \textbf{Min.} & \textbf{Med.} & \textbf{Max.} & & \textbf{Min.} & \textbf{Med.} & \textbf{Max.} & \\
    \midrule
    & \bug{1}  & \circom & 0 / 10 & & -- & \textbf{--} & -- & & -- & \textbf{--} & -- & \\
    & \bug{3} & \corset & 1 / 10 & & 6d 12h 32m & \textbf{6d 12h 32m} & 6d 12h 32m & & 273,132 & \textbf{273,132} & 273,132 & \\
    & \bug{7}  & \gnark & 10 / 10 & & 1h 22m & \textbf{4h 09m} & 23h 28m & & 1,236 & \textbf{3,857} & 21,865 & \\
    & \bug{9}  & \noir & 10 / 10 & & 10m & \textbf{1h 21m} & 4h 18m & & 528 & \textbf{3,865} & 12,742 & \\
    & \bug{10} & \noir & 9 / 10 & & 9h 00m & \textbf{3d 01h 23m} & 6d 19h 51m & & 15,098 & \textbf{123,584} & 279,353 & \\
    \midrule
    & \bug{4}  & \corset & 10 / 10 & & \($<$1\)m & \textbf{\($<$1\)m} & 3m & & 1 & \textbf{19} & 98 & \\
    & \bug{8}  & \gnark & 10 / 10 & & 14m & \textbf{2h 14m} & 3h 19m & & 206 & \textbf{2,041} & 3,228 & \\
    & \bug{11}  & \noir & 10 / 10 & & 15m & \textbf{2h 35m} & 11h 32m & & 378 & \textbf{4,399} & 19,012 & \\
    & \bug{12} & \noir & 0 / 10 & & -- & \textbf{--} & -- & & -- & \textbf{--} & -- & \\
    & \bug{13}  & \noir & 2 / 10 & & 3d 08h 12m & \textbf{5d 03h 53m} & 6d 23h 34m & & 142,009 & \textbf{221,024} & 300,039 & \\
    \bottomrule
  \end{tabularx}
\end{table*}

This section evaluates how quickly \tool rediscovers the bugs from
Table~\ref{tab:bugs}. We focus on bugs with developer-provided fixes.
This gives us a patch-based oracle for recognizing rediscoveries: a
test counts as a rediscovery of a target bug only if it exhibits the
target behavior on the vulnerable version and no longer exhibits that
behavior on the patched version. This filtering helps distinguish
rediscoveries of the target bug from other failures exposed during the
same campaign.

For each fixed bug, we ran ten campaigns with random seeds.  Each
campaign used one core and a seven-day timeout. For each campaign, we
measure the time until \tool first produces the same kind of
observation that originally exposed the target bug: an accepted
injected witness for bugs originally exposed through such witnesses,
and a crash or rejection for bugs originally exposed through
completeness failures. We report the number of successful seeds and
the minimum, median, and maximum values for both time to bug and
number of generated \il programs among successful campaigns.

Table~\ref{tab:ttb} summarizes the results. The full \tool
configuration rediscovers four of the five fixed bugs originally
exposed through accepted injected witnesses within the
timeout (above the horizontal line). Bugs~\bug{7} and~\bug{9} are
rediscovered in all ten seeds.  Bug~\bug{1} is not rediscovered by the
full configuration, but the ablated configuration in RQ5 rediscovers
it once.

The bugs exposed through accepted injected witnesses vary substantially
in how reliably they are rediscovered. Bugs~\bug{7} and~\bug{9} are
rediscovered in all ten campaigns, but for different reasons.
Bug~\bug{7} is exposed by a simple lying witness: successful tests use
divergence-point splicing and typically make one execution use scalar
zero. Bug~\bug{9} is an underconstrainedness issue in the lowering of
large ACIR constraints. Successful tests expose it with all splicing
strategies and many fault-input mutations, which explains why it is the
fastest soundness bug to find, with a median time to bug of 1h 21m.

The remaining bugs in this group are harder to
rediscover. Bug~\bug{10} is found in nine of ten seeds, with a median
time to bug of 3d 01h 23m.  This matches the structure of the bug:
\tool must generate a call to the relevant elliptic-curve routine,
sample inputs that produce different outputs, and splice the witness
so that the point coordinates remain consistent while the
\code{is\_infinite} flag and output come from the other
execution. Successful tests require shifted, windowed, or suffix
splicing rather than divergence-point splicing.

The bugs originally exposed through crashes or rejections (below the
horizontal line) are also rediscovered at different
rates. Bugs~\bug{4}, \bug{8}, and~\bug{11} are found in all ten
campaigns. Bug~\bug{4} is the easiest: the median campaign finds it in
less than one minute after generating 19 \il programs. Bugs~\bug{8}
and~\bug{11} take longer, with median times of 2h 14m and 2h 35m.

Bug~\bug{12} is not rediscovered in any campaign within the seven-day
timeout. It requires both the exact scalar value \(2^{128}\) and a
specific surrounding expression that causes \noir to represent one
part of the scalar decomposition as a variable and the other as a
constant.

Bug~\bug{13} is rediscovered in two of ten campaigns, with a median
time to bug of 5d 03h 53m. The bug incorrectly removes casts, but this
causes a crash only when the uncast value is later used by an
operation that requires the casted type, such as a bitwise operation
on a \code{u128}. In many generated programs, removing the cast
instead changes the computed value without causing a crash. The
advisory shows that the bug can also cause a soundness failure, but
adversarial witness injection does not expose this behavior. Cast
removal occurs before witness generation, so the initial executions
already reflect the miscompiled program. The oracle therefore lacks
the valid source-level executions needed to construct a lying witness
for this failure.

Overall, the full configuration rediscovers four of the five fixed bugs
originally exposed through accepted injected witnesses, while RQ5 shows
that adversarial witness injection can also rediscover bug~\bug{1}.
Rediscovery probability nevertheless varies substantially.
Bug~\bug{7} requires only divergence-point splicing, while bug~\bug{9}
occurs frequently in generated programs. Both are therefore
rediscovered consistently. Bugs requiring a rarer program shape, a
specific fault-input mutation, or careful intermediate-value splicing
are less reliable. The time-to-bug results measure the combined
effectiveness of three parts of \tool: generating a program that
reaches the relevant operation, sampling an input pair with different
outputs, and constructing an injected witness that preserves enough
structure to expose the missing or incorrect constraint.

%-------------------------------------------------------------------------------
\subsection{RQ4: Beyond \circuzz}
\label{sec:eval-circuzz}
%-------------------------------------------------------------------------------

\begin{table*}[t]
  \centering
  \setlength{\tabcolsep}{5pt}
  \caption{Time to bug for \circuzz. The \emph{Seeds} column reports
    the number of successful campaigns out of ten. \emph{Time} and
    \emph{\slantbox{\il} programs} report minimum/median/maximum values among
    successful campaigns; sub-minute times are shown as \($<$\)1m. The
    annotations beside \emph{Seeds} and the median columns show
    changes relative to the full \tool configuration in
    Table~\ref{tab:ttb}. Bugs above the horizontal line were
    originally exposed through accepted injected witnesses; bugs below
    it were originally exposed through crashes or rejections.}
  \label{tab:circuzz}
  \begin{tabularx}{\textwidth}{@{}X c r r X r r r r X r r r r X@{}}
    \toprule
    & \multirow{2.5}{*}{\textbf{Bug}} & \multicolumn{2}{c}{\multirow{2.5}{*}{\textbf{Seeds}}} & & \multicolumn{4}{c}{\textbf{Time to bug}} & & \multicolumn{4}{c}{\textbf{\il programs}} & \\
    \cmidrule(lr){6-9} \cmidrule(lr){11-14} & & & & & \textbf{Min.} & \multicolumn{2}{c}{\textbf{Med.}} & \textbf{Max.} & & \textbf{Min.} & \multicolumn{2}{c}{\textbf{Med.}} & \textbf{Max.} & \\
    \midrule
    & \bug{1}  & 1 / 10 & \positive{+1} & & 6d 17h 11m & \textbf{6d 17h 11m} & & 6d 17h 11m & & 193,858 & \textbf{193,858} & & 193,858 & \\
    & \bug{3}  & 0 / 10 & \negative{-1} & & -- & \textbf{--} & & -- & & -- & \textbf{--} & & -- & \\
    & \bug{7}  & 0 / 10 & \negative{-10} & & -- & \textbf{--} & & -- & & -- & \textbf{--} & & -- & \\
    & \bug{9}  & 0 / 10 & \negative{-10} & & -- & \textbf{--} & & -- & & -- & \textbf{--} & & -- & \\
    & \bug{10} & 0 / 10 & \negative{-9} & & -- & \textbf{--} & & -- & & -- & \textbf{--} & & -- & \\
    \midrule
    & \bug{4}  & 10 / 10 & \positive{=} & & \($<$1\)m & \textbf{1m} & \negative{+67\%} & 5m & & 1 & \textbf{11} & \positive{-42\%} & 24 & \\
    & \bug{8}  & 10 / 10 & \positive{=} & & 9m & \textbf{31m} & \positive{-77\%} & 5h 15m & & 140 & \textbf{611} & \positive{-70\%} & 6,014 & \\
    & \bug{11} & 10 / 10 & \positive{=} & & 9m & \textbf{7h 11m} & \negative{+178\%} & 1d 13h 45m & & 378 & \textbf{18,380} & \negative{+318\%} & 94,635 & \\
    & \bug{12} & 0 / 10 & \positive{=} & & -- & \textbf{--} & & -- & & -- & \textbf{--} & & -- & \\
    & \bug{13} & 10 / 10 & \positive{+8} & & 4m & \textbf{45m} & \positive{-99\%} & 6h 30m & & 152 & \textbf{1,851} & \positive{-99\%} & 16,102 & \\
    \bottomrule
  \end{tabularx}
\end{table*}

This section compares adversarial witness injection with \circuzz's
valid-execution oracle. The two oracles target different behaviors:
\circuzz checks valid, semantics-preserving executions, whereas
adversarial witness injection checks whether invalid witnesses are
rejected. We use the same fixed bugs, ten-seed campaign setup,
seven-day timeout, and patch-based rediscovery criterion as in
Section~\ref{sec:eval-ttb}. The only change is the oracle: instead of
injecting adversarial witnesses, these campaigns use \circuzz's
metamorphic-testing oracle.

Table~\ref{tab:circuzz} reports the \circuzz results. The upper part
contains bugs originally exposed through accepted injected witnesses,
while the lower part contains bugs originally exposed through crashes
or rejections. The table also compares the number of successful
campaigns and, where both configurations succeed, the median time and
number of generated \il programs with the full \tool results from
Table~\ref{tab:ttb}.

The main result is that \circuzz does not expose any of the soundness
failures revealed by accepted injected witnesses. It does not
rediscover bugs~\bug{3}, \bug{7}, \bug{9}, or~\bug{10}. These are
underconstrainedness bugs for which valid executions provide no
disagreement for a metamorphic oracle to observe.

Bug~\bug{1} is different because the same sign error can both accept
invalid witnesses and reject valid ones. \circuzz rediscovers the root
cause once through a completeness failure, but does not expose its
soundness impact. Conversely, the ablated \tool configuration in RQ5
rediscovers the soundness failure once. The single rediscovery by each
configuration suggests that the bug is difficult to reach within the
seven-day budget and that rediscovery is highly dependent on the
random seed.

For bugs originally exposed through completeness failures, the picture
is different. \circuzz rediscovers bugs~\bug{4}, \bug{8}, \bug{11},
and~\bug{13} in all ten campaigns. The corresponding \tool campaigns
also rediscover bugs~\bug{4}, \bug{8}, and~\bug{11} in all ten seeds,
but their relative efficiency varies. For bug~\bug{4}, \circuzz uses
42\% fewer \il programs, although its median time is 67\% higher. For
bug~\bug{8}, it reduces the median time by 77\% and the number of
generated \il programs by 70\%. For bug~\bug{11}, it increases them by
178\% and 318\%, respectively.

Bug~\bug{13} differs primarily in reliability. \circuzz rediscovers it
in all ten campaigns, eight more than \tool. As discussed in
Section~\ref{sec:eval-ttb}, the cast-removal bug can affect valid
executions, making it well matched to \circuzz's oracle. The table also
shows 99\% lower median time and program count for \circuzz, but these
medians cover ten successful campaigns for \circuzz and only two for
\tool. They therefore do not support a direct comparison of speed.

The only fixed bug originally exposed through a completeness failure
that \circuzz does not rediscover is bug~\bug{12}. As discussed in
Section~\ref{sec:eval-ttb}, its rare trigger also prevents \tool from
rediscovering it within the seven-day timeout.

Overall, adversarial witness injection exposes soundness failures that
the valid-execution oracle does not. \circuzz produces no rediscoveries
for bugs~\bug{3}, \bug{7}, \bug{9}, and~\bug{10}, which are pure
underconstrainedness bugs. It rediscovers bug~\bug{1} through a
completeness failure, but does not expose its soundness impact. For
bugs~\bug{4}, \bug{8}, and~\bug{11}, whose completeness failures are
visible during valid executions, both approaches rediscover the bugs
reliably, with no consistent efficiency advantage.

%-------------------------------------------------------------------------------
\subsection{RQ5: Effect of \il Extensions}
\label{sec:eval-ablation}
%-------------------------------------------------------------------------------

\begin{table*}[t]
  \centering
  \setlength{\tabcolsep}{4pt}
  \caption{Effect of disabling the new \il program-generation
    features. Each entry reports the number of successful ablated
    campaigns out of ten. \emph{Time} and \emph{\slantbox{\il}
    programs} report medians over successful ablated campaigns;
    sub-minute median times are shown as \($<$\)1m. The annotations
    beside the seed counts and medians show changes relative to the
    corresponding full configuration in Tables~\ref{tab:ttb}
    and~\ref{tab:circuzz}.}
  \label{tab:ablation}
  \begin{tabularx}{\textwidth}{@{} X c X r r r r r r X r r r r r r X @{}}
    \toprule
    & \multirow{2.5}{*}{\textbf{Bug}} & & \multicolumn{6}{c}{\textbf{\tool ablated}} & & \multicolumn{6}{c}{\textbf{\circuzz ablated}} & \\
    \cmidrule(lr){4-9} \cmidrule(lr){11-16}
    & & & \multicolumn{2}{c}{\multirow{2}{*}{\textbf{Seeds}}} & \multicolumn{2}{c}{\multirow{2}{*}{\textbf{Time}}} & \multicolumn{2}{c}{\textbf{\il}} & & \multicolumn{2}{c}{\multirow{2}{*}{\textbf{Seeds}}} & \multicolumn{2}{c}{\multirow{2}{*}{\textbf{Time}}} & \multicolumn{2}{c}{\textbf{\il}} & \\
    & & & \multicolumn{2}{c}{} & \multicolumn{2}{c}{} & \multicolumn{2}{c}{\textbf{programs}} & & \multicolumn{2}{c}{} & \multicolumn{2}{c}{} & \multicolumn{2}{c}{\textbf{programs}} & \\
    \midrule
    & \bug{1}  & & 1 / 10 & \positive{+1} & 6d 20h 45m & & 53,133 & & & 1 / 10 & \positive{=} & 2d 16h 04m & \positive{-60\%} & 80,804 & \positive{-58\%} & \\
    & \bug{3}  & & 1 / 10 & \positive{=} & 4d 02h 55m & \positive{-37\%} & 186,141 & \positive{-32\%} & & 0 / 10 & \positive{=} & -- & & -- & & \\
    & \bug{7}  & & 0 / 10 & \negative{-10} & -- & & -- & & & 0 / 10 & \positive{=} & -- & & -- & & \\
    & \bug{9}  & & 10 / 10 & \positive{=} & 16m & \positive{-80\%} & 887 & \positive{-77\%} & & 0 / 10 & \positive{=} & -- & & -- & & \\
    & \bug{10} & & 0 / 10 & \negative{-9} & -- & & -- & & & 0 / 10 & \positive{=} & -- & & -- & & \\
    \midrule
    & \bug{4}  & & 10 / 10 & \positive{=} & \($<$1\)m & \positive{-50\%} & 21 & \negative{+11\%} & & 10 / 10 & \positive{=} & \($<$1\)m & \positive{-13\%} & 11 & \positive{=} & \\
    & \bug{8}  & & 0 / 10 & \negative{-10} & -- & & -- & & & 0 / 10 & \negative{-10} & -- & & -- & & \\
    & \bug{11} & & 0 / 10 & \negative{-10} & -- & & -- & & & 0 / 10 & \negative{-10} & -- & & -- & & \\
    & \bug{12} & & 0 / 10 & \positive{=} & -- & & -- & & & 0 / 10 & \positive{=} & -- & & -- & & \\
    & \bug{13} & & 0 / 10 & \negative{-2} & -- & & -- & & & 10 / 10 & \positive{=} & 7h 58m & \negative{+962\%} & 20,171 & \negative{+990\%} & \\
    \bottomrule
  \end{tabularx}
\end{table*}

This section evaluates how much the new \il program-generation
features contribute to the results. We repeated the time-to-bug
experiments from Section~\ref{sec:eval-ttb} with an ablated
configuration that disables arrays, let expressions, and function
specifications. We evaluated the ablation with both adversarial
witness injection and \circuzz's metamorphic-testing oracle. This
separates the effect of richer program generation from that of the
oracle.

Table~\ref{tab:ablation} reports the ablated results. Each entry
reports how many of ten random seeds rediscovered the target bug. For
successful campaigns, the table also reports the median time to bug
and median number of generated \il programs. The annotations compare
the seed counts and medians with the corresponding full configurations
from Tables~\ref{tab:ttb} and~\ref{tab:circuzz}.

The clearest ablation effects occur for bugs that require operations
introduced through function specifications. As discussed in
Section~\ref{sec:eval-characteristics}, bugs~\bug{7} and~\bug{10}
require generated programs to call elliptic-curve library routines,
while bug~\bug{8} involves lookup functionality. The ablated \tool
configuration cannot generate the relevant calls and therefore does not
rediscover any of these bugs. Bug~\bug{12} also requires an
elliptic-curve scalar operation. The ablated configuration cannot reach
this operation at all, while the full configuration also fails to
rediscover the bug because of its rare trigger.

The remaining bugs originally exposed through accepted injected
witnesses do not depend on standard-library calls, and their ablation
results are more mixed. Bug~\bug{9} is rediscovered in all ten
campaigns with both configurations. However, the ablated configuration
reduces the median time by 80\% and the number of generated \il programs
by 77\%. A plausible explanation is that disabling the new features
produces simpler generated programs in which the constraint structure
needed to expose this general ACIR-lowering bug occurs more frequently.

Bug~\bug{1} is not rediscovered by the full \tool configuration but
is rediscovered once by the ablated configuration. This isolated
success does not indicate that the ablation improves effectiveness:
the bug is difficult to reach, and its rediscovery appears highly
dependent on the random seed. Bug~\bug{3} is rediscovered once by each
configuration. Although the successful ablated campaign is 37\%
faster and generates 32\% fewer \il programs, one successful campaign
per configuration does not support a meaningful efficiency comparison.

Among the bugs originally exposed through crashes or rejections, only
bug~\bug{4} remains reliably rediscoverable by the ablated \tool
configuration. It is found in all ten campaigns, with a median time
below one minute. This is expected: the \corset checker bug does not
require any of the added language features.

Bug~\bug{11} is not rediscovered by the ablated configuration. As
discussed in Section~\ref{sec:eval-characteristics}, the bug is
triggered by redundant target-level assertions that can arise
from richer expressions and standard-library calls. Without the new
\il features, this pattern is effectively unreachable.

Bug~\bug{13} is rediscovered twice by the full configuration but not
by the ablated configuration. The full configuration includes custom
function specifications whose translations introduce additional casts,
substantially increasing the opportunities to trigger this cast-removal
bug.

The ablated \circuzz results show that the new \il features also help
metamorphic testing. For the bugs originally exposed through accepted
injected witnesses, neither \circuzz configuration exposes a
soundness failure. Bug~\bug{1} remains the exception at the level of
the root cause: both configurations rediscover it once through a
completeness failure, but neither exposes its soundness impact.

For bugs originally exposed through completeness failures,
bug~\bug{4} remains easy to rediscover, while bugs~\bug{8}
and~\bug{11} disappear when the new \il features are disabled.
Bug~\bug{12} is not rediscovered by either configuration.

Bug~\bug{13} remains reliable in the ablated \circuzz configuration,
but its median time increases from 45 minutes to almost eight hours,
and its median program count increases by 990\%. As with the \tool
ablation, this is consistent with the new function specifications
creating more opportunities to trigger the bug.

Overall, the ablation shows that richer program generation expands the
bugs reachable by both oracles. Function specifications are essential
for reaching bugs in elliptic-curve and lookup functionality and
create additional opportunities to trigger cast-related bugs. The new
features also appear important for generating the pattern behind
bug~\bug{11}. Conversely, simpler programs can make lower-level bugs
that do not require the new features faster to rediscover, as
illustrated by bug~\bug{9}. The extensions therefore broaden the
tested functionality, although they do not uniformly reduce time to
bug.

%-------------------------------------------------------------------------------
\subsection{Threats to Validity}
\label{sec:eval-threats}
%-------------------------------------------------------------------------------

Our results depend on the ZK toolchains under test, their versions and
configurations, and the \tool configuration. We reduce this threat by
evaluating four toolchains that differ in their source languages,
witness representations, and proving or checking workflows. We also
compare multiple configurations: adversarial witness injection,
\circuzz's metamorphic-testing oracle, and ablated variants without the
new \il features. Because the ablation disables arrays, let
expressions, and function specifications together, it measures their
combined effect. We attribute individual results to function
specifications only when the relevant operations cannot be generated
without them.

Fuzzing is randomized, and the time needed to rediscover a bug can
vary substantially across seeds. We account for this by running ten
independent campaigns for each fixed-bug experiment and by reporting
the number of successful seeds and the minimum, median, and maximum
time to bug and number of generated \il programs. These values are
computed only over successful campaigns. Comparisons are therefore
most informative when configurations have similar success counts; in
other cases, we treat the reported values descriptively rather than as
evidence of relative efficiency.

The adversarial-witness oracle assumes deterministic source semantics
and that the generated program is intended to constrain its public
input-output relation. If a source construct intentionally leaves
values unconstrained, acceptance of an injected witness may be expected
rather than indicate a toolchain bug. We therefore generate programs
with public inputs and outputs, exclude features such as \circom's
unconstrained assignments, and manually inspect reported soundness
bugs.

False negatives remain possible. Coverage depends on the generated
programs, sampled inputs, witness-splicing strategies, and
target-specific operations supported by the current implementation.
Bugs that require rare program shapes, specific constants,
standard-library routines that \tool cannot yet generate, or
combinations of intermediate values not produced by the supported
witness-splicing strategies may not be exposed within the campaign
timeout.

%%% Local Variables:
%%% mode: latex
%%% TeX-master: "main"
%%% End:

%% file: related.tex
%!TEX root = main.tex

%-------------------------------------------------------------------------------
\section{Related Work}
\label{sec:related}
%-------------------------------------------------------------------------------

Related work spans compiler testing, testing ZK systems, testing and
analyzing ZK circuits, and formal verification of ZK compilers.

\textbf{Testing compilers.}
Compiler testing commonly applies differential or metamorphic oracles
to generated programs or to mutations of existing programs
~\cite{ChenPatra2020}. These techniques have been applied across
compiler domains, including
conventional~\cite{YangChen2011,LeAfshari2014},
graphics-shader~\cite{DonaldsonEvrard2017}, and MPC
compilers~\cite{WatzingerWuestholz2026}.

\textbf{Testing ZK systems.}
Compiler-testing techniques have also been adapted to ZK systems,
whose toolchains combine DSL compilers, witness generators, provers,
and verifiers. \circuzz~\cite{HochrainerIsychev2025} applies
metamorphic testing end to end to ZK DSL toolchains, while
MTZK~\cite{XiaoLiu2025} focuses on ZK compilers. Both test
semantics-preserving executions and can expose bugs whose effects are
observable during such executions, but do not directly test whether
invalid witnesses are rejected. Chaliasos et al.\ identify constructing
invalid witnesses without violating unrelated constraints as an open
challenge~\cite{ChaliasosAlFath2025}. \tool addresses this challenge by
splicing witnesses from valid executions with different outputs,
yielding witnesses that cannot correspond to valid source-level
executions.

\arguzz~\cite{HochrainerWuestholz2026} simulates a malicious prover,
combining metamorphic testing with fault injection inside zkVM
provers. \tool instead constructs invalid witnesses from valid
executions and injects them into ZK DSL
toolchains. SNARKProbe~\cite{FanXu2024} combines dynamic analysis,
fuzzing, and SMT solving to test cryptographic libraries and prover
implementations. \tool instead tests generated DSL programs through
the complete toolchain.

\textbf{Testing and analyzing ZK circuits.}
Kolozyan et al.~\cite{KolozyanSorger2026} systematize security
tools for zero-knowledge systems and evaluate six automated tools on
70 real-world vulnerabilities. They find that current tools focus
mainly on \circom and nondeterminism, with weaker support for semantic
mismatches and newer DSLs. Their study characterizes the broader
tooling landscape; \tool contributes a new testing oracle for complete
ZK DSL toolchains.

A complementary line of work targets vulnerabilities in ZK application
circuits rather than toolchain infrastructure. Chaliasos et
al.\ distinguish circuit-level bugs, often introduced by circuit
developers, from infrastructure bugs in compilers and proof-system
implementations~\cite{ChaliasosErnstberger2024}. Circuit-level
approaches use fuzzing~\cite{FuTan2026,TakahashiKim2026}, SMT-based
analysis~\cite{PailoorChen2023,SoureshjaniHallAndersen2023,%
IsabelRodriguezNunez2024,JiangPeng2025,StephensPailoor2025},
and static or value-inference
analyses~\cite{WenStephens2024,JiangQin2025,%
KolozyanVandenbogaerde2026}. Other work uses refinement types to
certify properties of circuits generated from a dedicated
language~\cite{LiuKretz2024}.

These techniques analyze source circuits, compiled constraint systems,
or the consistency between witness computation and constraints. \tool
instead tests whether the complete toolchain accepts invalid
witnesses. The two lines of work are complementary: circuit-level
analyses may detect bugs already present in the source circuit or
compiled constraint system, including standard-library bugs that
affect the generated constraints. They do not target bugs introduced
later in the toolchain. For example, in bug~\bug{1}, the generated
R1CS is correct but its conversion inside the prover introduces a sign
error; in bug~\bug{9}, the ACIR is correct but the prover backend
drops constraints during lowering.

\textbf{Formal verification of ZK compilers.}
Formal verification has produced strong correctness guarantees for
general-purpose compilers, most notably
CompCert~\cite{Leroy2009}. Similar techniques have been applied to ZK
compilation. ZKCrypt compiles high-level proof goals into
implementations of zero-knowledge protocols and formally verifies the
security of the generated
implementations~\cite{AlmeidaBarbosa2012}. Ozdemir et al.\ partially
verify the field-blasting pass of the CirC compiler using bounded SMT
verification~\cite{OzdemirWahby2025}. These approaches provide
stronger guarantees for the verified components, whereas \tool tests
complete toolchains, including unverified components and their
integration.

%%% Local Variables:
%%% mode: latex
%%% TeX-master: "main"
%%% End:

%% file: conclusion.tex
%!TEX root = main.tex

%-------------------------------------------------------------------------------
\section{Conclusion}
\label{sec:conclusion}
%-------------------------------------------------------------------------------

ZK DSL toolchains must reject witnesses that violate source-program
semantics. We introduced adversarial witness injection, which constructs
invalid witnesses by splicing executions with different outputs.
Controlled divergence and splicing strategies expose missing
constraints, while function specifications reach standard-library code.

Across four toolchains, \tool found 13 bugs, seven with
soundness impact; six were exposed by accepted injected
witnesses. These bugs span compilers, witness generation,
constraint lowering, standard libraries, and prover backends.
Under the same budget, \circuzz exposed none of the
soundness failures but remained effective for completeness
bugs.

Adversarial witness injection turns the oracle problem into a
construction problem: \tool constructs witnesses known to violate
source-program semantics and tests whether the toolchain rejects them.
It complements valid-execution testing by testing not only what systems
accept, but also what they must reject.

%%% Local Variables:
%%% mode: latex
%%% TeX-master: "main"
%%% End: